\documentclass[paper]{JHEP3}
\usepackage{epsfig}
\def\icms{cm$^{-2}$s$^{-1}$}
\def\inb{nb$^{-1}$}
\def\lum{{\cal{L}}}
\def \lsim{\mathrel{\vcenter
     {\hbox{$<$}\nointerlineskip\hbox{$\sim$}}}}

\def    \be             {\begin{equation}}
\def    \ee             {\end{equation}}
\def    \ba             {\begin{eqnarray}}
\def    \ea             {\end{eqnarray}}

\def    \=              {\;=\;}
\def    \frac           #1#2{{#1 \over #2}}
\def    \mh             {\mbox{$m_H$}}

\newcommand{\gev}{\mbox{GeV}}

\def\met{$\rlap{\kern.2em/}E_T$}
\title{ LHC physics: the first one--two year(s)... 
\footnote{Presented at the 2nd Italian Workshop on the Physics
  of Atlas and CMS, Napoli, October 13-15 2004.}}
\vfill                                                       
\author{
Fabiola Gianotti and Michelangelo Mangano        \\
{\em CERN, PH Department, Geneva, Switzerland} \\}
\abstract{
  We discuss the strategy to commission the LHC
 experiments and understand standard physics at $\sqrt{s}=14$~TeV
 before data taking starts and in the early phases of the LHC 
  operation. 
  In particular, we review the various steps 
  needed to understand and calibrate the ATLAS
 and CMS detectors, from construction quality
 checks, to beam tests, to cosmics runs, to first collisions.
  We also review  the preparation and tuning 
 of Monte Carlo tools, and present a few examples of physics goals 
 for integrated luminosities of up to a few fb$^{-1}$. 
}
\preprint{CERN-PH-TH/2005-072\\hep-ph/0504221 }
\begin{document}

\section{Introduction}
When the LHC will start providing data to the experiments,
unprecedented opportunities to explore the frontier of high energy
physics as we know it today will suddenly become
available\cite{Mangano:2002jw}.  It will
take some time before the accelerator ramps up in
luminosity and the Collaborations debug and understand their
detectors. Nevertheless it is crucial to realise that possible new
exotic phenomena could have cross sections so large, and topologies so
striking, that even a limited amount of collected data and a
non-ultimate detector performance could lead to exciting
results. Readiness to capture these opportunities is a must.
Whether nature is kind to us and is preparing a sweet welcome
to the TeV energy frontier, or whether we shall have to sweat through
years of hard work before the total luminosity and the detector
performance allow us to establish the existence of new phenomena, the
preparation for the first phase of data taking, which includes both
the definition of the strategies for the commissioning of the
detectors and triggers and of the physics analyses, is therefore a
task of great priority. 

 This presentation offers an elementary review
of the physics
landscape which ATLAS and CMS could be exposed to in the early days of
running, and discusses the efforts which are taking place to ensure a prompt
exploitation of the new data.  %

\section{Physics opportunities at the beginning}
In the Fall of 1982 the first extended physics run of UA1 and UA2 took
 place at the CERN $S\bar{p}pS$ collider, at $\sqrt{S}=546$~GeV. 
The maximum luminosity was a mere
$\lum =5\times 10^{28}$cm$^{-2}$s$^{-1}$ ($\sim 1\%$ of the asymptotic one), 
with a total sample
of  $\sim 20$\inb\  integrated over 30 days. The outcome of this run was
 nevertheless a big hit: the discovery of $W$ and $Z$ bosons,
 established after only few months of hard work\cite{Arnison:rp}. 
However tough the new
 experimental environment, however weak the understanding of physics
 at those unprecedented energies (jets had only been first observed
 few months earlier, after a yet shorter pilot run in
 1981~\cite{Banner:1982kt}),  the
 rate and features of the signal were such that it could not be
 missed. This was not a surprise: a key
 role was played by the energy, which being almost a factor of 10 larger than
 the previous frontier, the ISR, opened up the required phase
 space. 

Few years later, in the Summer of 1987, the first physics
 run for CDF at the Fermilab Tevatron collider ($\sqrt{S}=1.8$~Tev)
took place, again at
 $\lum=5\times 10^{28}$\icms\ ($\sim 1\%$ of the
 design value), with a total of $\sim 20$\inb.
Nothing new emerged from this run, and it did take some time after
 that before CDF could start exploring truly new territory. The reason
 is that the jump in energy by a factor of 3 was not large enough to
 compensate for the integrated luminosity already accumulated by
 UA1 and UA2. Over 100\inb\ would have been necessary to improve on,
 say, the top quark search, as the production cross section at the
 Tevatron was ``only'' a factor of 10-20 larger than at CERN, in the
 relevant range of masses. 

\begin{figure}[t]
\begin{center}
\epsfig{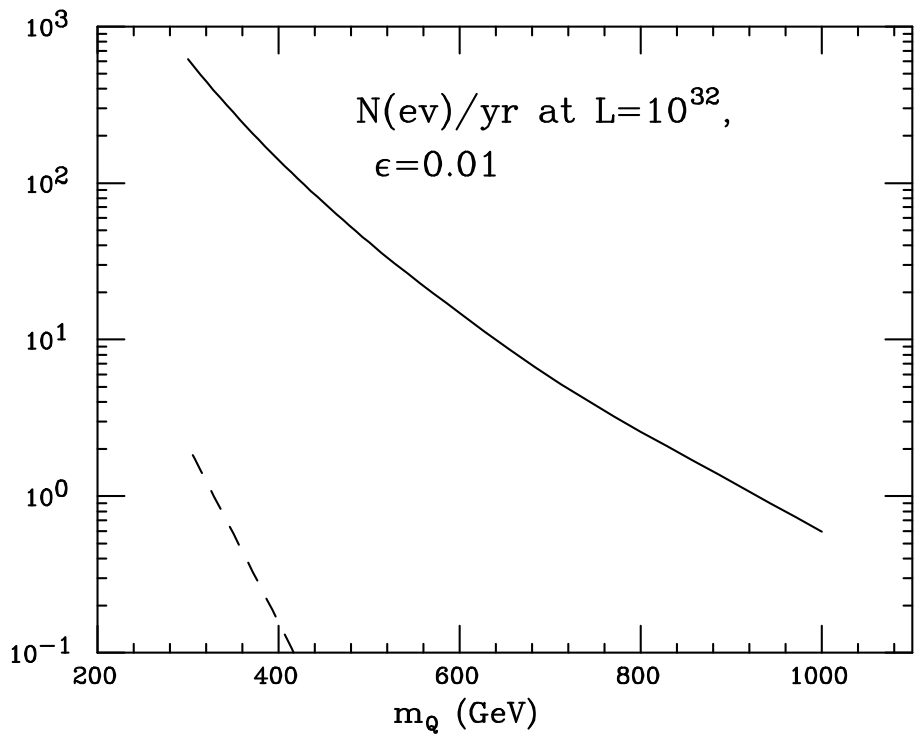}
\caption{\it Production rates for heavy quark pairs, as a function of
 the quark mass, at the Tevatron
 (dashed) and at the LHC (solid), during one year of data taking at 
$10^{32}$\icms, and assuming a detection efficiency of
 1\%. \label{fig:topxs}}	 
\end{center}
\end{figure}

When the LHC will start, the situation will be much more like that at the
time of the $S\bar{p}pS$ turn on. In spite of the
multi-fb$^{-1}$ luminosity which we expect CDF and D0 to collect by
that time,
rates for new particles (heavy quarks, gluinos, new gauge bosons, etc.)
with mass beyond the discovery reach of the Tevatron will allow their
 abundant production already with typical start-up
luminosities of 1\% of the design, namely $\lum \sim
10^{31-32}$\icms. This is clearly shown in fig.~\ref{fig:topxs}, which
plots the production rate for pairs of new heavy quarks 
(already at the rather low mass of the top quark the
rate at the LHC is over 100 times larger than at the Tevatron!).
Knowing that cross sections for gluinos are typically one order of
magnitude larger than for quarks of equal mass, this figure gives also
a clear picture of the immense Supersymmetry (SUSY) discovery 
potential of early LHC data!

So, we have phase-space, we have large rates for new physics. 
But should we seriously expect
something to show up at the LHC energy scale and at and luminosities 
reachable early on? The Tevatron and LEP's heritage is a
strong confirmation of the Standard Model (SM), and at the same time an apparent
paradox\cite{Barbieri:2004ii}, illustrated in the following paragraphs.
Electroweak (EW) precision tests and the value of the top mass are
consistent with, and require, a rather light Higgs mass: $m_H=117{
  +45\atop -68}$ GeV; 
EW radiative corrections in the SM, integrated up to a scale $\Lambda$,  
shift the bare value of $m_H$ by:
\be
\delta m_H^2 = \frac{6G_F \Lambda^2}{\sqrt{2}\pi^2} \; (m_t^2 -
\frac{1}{2} m_W^2 - \frac{1}{4} m_Z^2 - \frac{1}{4} m_H^2)
\sim (115\;\mbox{GeV})^2 \; \left( \frac{\Lambda}{400\;\mbox{GeV}}
\right)^2. \label{eq:mh}
\ee
%\begin{eqnarray}
%\delta m_H^2 &=& \frac{6G_F \Lambda^2}{\sqrt{2}\pi^2} \; (m_t^2 -
%\frac{1}{2} m_W^2 - \frac{1}{4} m_Z^2 - \frac{1}{4} m_H^2) \nonumber
%\\
%&\sim& (115\;\mbox{GeV})^2 \; \left( \frac{\Lambda}{400\;\mbox{GeV}}
%\right)^2. \label{eq:mh}
%\end{eqnarray}
The integration in principle can extend up to very large values of
$\Lambda$, where new particles may appear, changing eq.~(\ref{eq:mh}).
As $\Lambda$ gets significantly larger than 400~GeV, however, the
presence of a counterterm (CT) should be assumed, to ensure that the
overall value of \mh\ is consistent with its bounds. This
CT can be interpreted as a 
low-energy manifestation of the physical mechanisms which, at some scale
$\overline{\Lambda}$,  modify eq.~(\ref{eq:mh}).
 Ensuring that the residual of the cancellation between
eq.~(\ref{eq:mh}) and the CT is in the 100~\gev\ range, however,
forces a fine tuning which becomes more and more unbelievable as 
$\overline{\Lambda}$ grows. Assuming that no new physics appears
before the GUT scale of $10^{16}$~\gev\ would lead to a level of fine
tuning of $10^{-28}$! By and large theorists believe that this is
unlikely enough to call for the existence of new physics at scales in
the range of 1--few TeV, so as to maintain the fine tuning level to
within ${\cal O}(10^{-3})$. This belief however clashes (and this is
where the paradox arises) with the staggering agreement between EW
data and the SM. The inclusion of generic new physics, parameterized in
terms of low-energy effective couplings between the SM particles, and
the analysis of the effects induced on EW observables, set lower limits to
the scale   $\overline{\Lambda}$ in the range of
5-10~TeV\cite{Barbieri:1999tm}, at the extreme 
limit of the fine-tuning window.  The
solution to the paradox could only be obtained with new physics which
cancels the large radiative contributions to \mh\
and, at the same time, manages to leave all other EW parameters and observables
unaffected. SUSY provides one such example! The cancellation
of large loop effects between SM particles and their SUSY partners
modifies eq.~(\ref{eq:mh}) and leads to an upper limit on \mh, given
in a simplified approximation here:
\be \label{eq:mhsusy}
m_H^2 \lsim m_Z^2 +\frac{3G_F \, m_t^4}{\sqrt{2}\pi^2} \log\left(
\frac{m_{\tilde{t}}^2}{m_t^2} \right )
\ee
where $m_{\tilde{t}}^2$ is the average squared mass of the two stop states.
At the same time, the structure of the theory is such that indeed
generic choices of the SUSY parameters, consistent with current experimental
limits on new particles, lead to negligible effects in the EW observables.
In the minimal realization of SUSY (MSSM), 
when eq.~(\ref{eq:mhsusy}) is improved with 2-loop and
non-logarithmic corrections, the experimental limit on \mh\ pushes however the
scale of SUSY in the multi-TeV domain. Once again this is at the edge of being
acceptable as a ``natural'' solution to the fine-tuning problem,
and for many theorists the room left for SUSY is becoming too tight.
As a result, new scenarios for EW symmetry breaking, particularly some
where the upper limits on the Higgs mass are looser compared to the
MSSM, have been proposed (as reviewed in\cite{Barbieri:2004ii}). 

While
these alternative scenarios could take much longer to be
identified experimentally, the SUSY framework provides a strong and
appealing physics case for possible early discovery, and therefore
should be given maximum priority in the planning for the first data
analyses. 
SUSY is in fact expected to manifest itself with abundant and striking
signals, such as the production of multijets with large missing
transverse
energy (\met), multileptons (possibly same-charge), or prompt photons with
large \met. Because of rates, background levels, and nature of the
observables, searches for SUSY are expected to be  less demanding
from the experimental point of view than the quest for the Higgs in
the $\mh<140$~\gev\ range.
 In addition, SUSY provides a natural candidate for dark
matter, namely the lightest neutralino $\chi^0_1$,
the neutral SUSY partner of the photon and $Z$. Proving the
direct link between dark matter and SUSY would be, perhaps even more than
the Higgs discovery, the flagship achievement of the LHC! Last but not
least, an
early detection of SUSY could immediately provide clear directions to
the field of experimental high-enegy physics, and allow a robust
planning for future facilities.

\section{Machine start-up scenario}

   According to the present LHC schedule (see\cite{rossi} for more details),
  the machine will be cooled down in Spring 2007, 
  and will then be commissioned for a few months 
  starting with single beams. A first run with colliding beams is
  expected in the second half of 2007, and will likely
 be followed by a shut-down of a few months, and then by 
 a seven-month physics run in 2008 at instantaneous luminosities  of up to 
 $2\times 10^{33}$~cm$^{-2}$~s$^{-1}$.

    There are several uncertainties on this plan (in particular
   because of the recent problems with the production of the cryogenic
   line) and on how the machine  commissioning and performance will 
  actually evolve. Therefore  we assume here that the
  integrated luminosity collected by the end of 2008 will range 
  between a very modest 100~pb$^{-1}$ per experiment and a 
  very ambitious 10~fb$^{-1}$
  per experiment, and we discuss the LHC physics potential for this range.

\section{Initial detectors and initial performance\label{initial}}

    The first question to address is which detectors will
   be available at the beginning. 
  Indeed, because of missing resources, and in some cases of 
 construction delays, several components of ATLAS and
CMS will not be complete at the beginning of
data taking. ATLAS will start  with
two pixel layers (instead of three) and without 
Transition Radiation Tracker in the region $2<|\eta|<2.4$. 
 CMS will start without muon trigger chambers (RPC) in the  
 region $1.6<|\eta|<2.1$ and without the fourth layer of the end-cap
 muon chambers. Furthermore, the CMS 
 end-cap electromagnetic calorimeter and pixel detector will be installed 
 during the shut-down period after the 2007 run. 
 In addition, in both experiments part of the high-level
 trigger and data acquisition processors will be deferred, with the
 consequence 
 that the output rate of the level-1 trigger will be limited to
 50~kHz (instead of 100~kHz) in CMS and to 35~kHz (instead of 75~kHz)
 in ATLAS. 
 
 The impact of this staging on physics will be significant but not
 dramatic. The main loss is a descoped $B$-physics programme because, 
 due to the reduced level-1 bandwidth, the thresholds of the
 single-muon triggers will have to be raised from a few GeV
 (as originally chosen to address $B$-physics studies) to 
 $p_T$=14-20~GeV.

\begin{table}[t]
\centering
\caption{\it Examples of expected detector performance for ATLAS and
 CMS at the time of the LHC start-up, and 
 of physics samples which will be used to
 improve this performance.} 
\vskip 0.1 in 
\begin{tabular}{|l|c|c|c|}
\hline 
  & expected performance &
data samples (examples)  \\
 &  on ``day 1" & to improve the performance  \\
\hline\hline
ECAL uniformity & $\sim$1\% ($\sim$4\%) in ATLAS (CMS) 
& minimum-bias, $Z\to ee$\\
electron energy scale & 1-2\% & $Z\to ee$ \\
HCAL uniformity & 2-3\% & single pions, QCD jets \\
jet energy scale & $\leq$10\% 
& $Z (\to\ell\ell)$+jet, $W\to jj$ in $t\overline{t}$ events \\
tracker alignment & 20-200~$\mu$m in $R\phi$ 
& generic tracks, isolated $\mu$, $Z\to\mu\mu$ \\
\hline
\end{tabular}
\label{tab_perf}
\end{table}
 
      The second question concerns the detector
     performance to be expected on ``day 1", i.e. at the moment
     when data taking starts.  Some predictions,  based 
  on construction quality checks, on the known precision of the 
  hardware calibration and alignment systems, 
  on test-beam measurements and 
 on simulation studies, are given in tab.~\ref{tab_perf}
 for illustration.  
  The initial uniformity of the electromagnetic
 calorimeters (ECAL) should be at the level of 1\% for the 
 ATLAS liquid-argon calorimeter and 4\% for the CMS crystals, where
 the difference comes from the different techniques and from
 the limited time available for test-beam measurements  in CMS. 
  Prior to data taking, the jet energy
 scale may be established to about 10\% from a combination of 
 test-beam measurements and simulation studies. 
  The tracker alignment in the transverse plane is expected to be
 known at the level of 20~$\mu$m in the best case   
  from  surveys, from the hardware alignment systems, and possibly
  from some studies with cosmic muons and beam halo events.
  
\begin{figure}[t]
\begin{center}
\epsfig{file=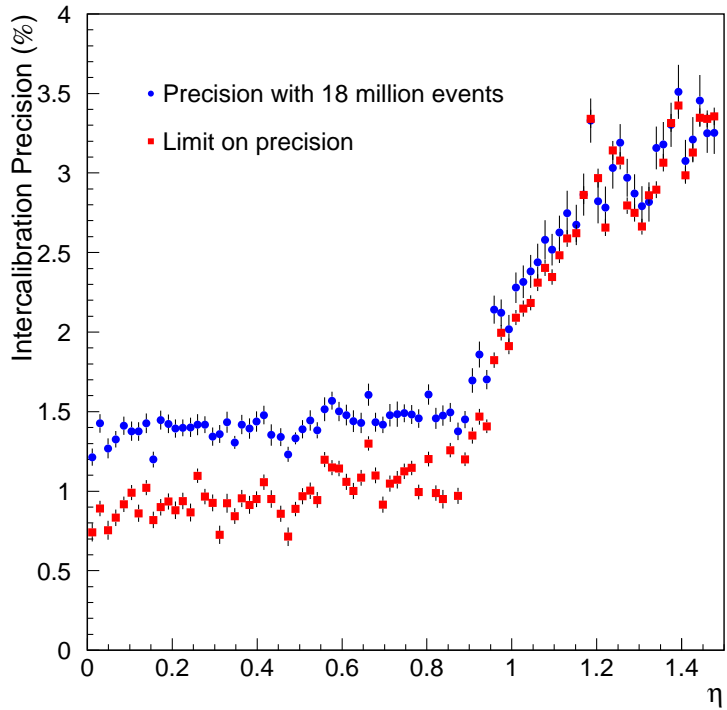,width=0.7\textwidth}
\caption{\it Inter-calibration precision of the CMS electromagnetic calorimeter
 achievable with 18 million minimum-bias
 events\cite{ref_cmsecal}, as a function of rapidity (dots). 
 The squares show the limit coming from the 
 non-uniformity of the upstream material.\label{cms_ecal}}	 
\end{center}
\end{figure}

   This performance should be significantly improved 
   as soon as the first data
  will be available (see last column in tab.~\ref{tab_perf}) and,
    thanks to the huge event rates expected at the LHC, the ultimate 
   statistical precision should be achieved after a 
   few days/weeks of data taking.  
    Then the painful battle with the systematic uncertainties will start.
   This is illustrated in fig.~\ref{cms_ecal} which shows that, by 
   measuring the energy flow in 
   about 18~million minimum-bias events (which can be collected in principle
   in a few hours of data taking), the non-uniformity of the CMS ECAL
   should be reduced from the initial 4\% to about 1.5\%
   in the barrel region. Therefore 
   the systematic limit coming from the 
   non-uniformity of the upstream tracker material  
   will be hit very quickly. 

%    Hopefully, with the help of a large variety of
%   high-statistics physics samples, close-to-final performance
%   for many important issues will be achieved 
%   after about one year of operation  (e.g. 
%   tracker alignment to $<5\ \mu$m, ECAL uniformity to $\sim$0.5\%, 
%   jet energy scale to $\sim$1-2\%). 

\section{Strategy to achieve the goal detector performance\label{larg}}

   Are the performance expectations 
  presented in the previous section realistic?
  This is discussed below with the help of a concrete example. 

   The ATLAS and CMS detectors have been subject to stringent 
  requirements and detailed quality controls 
  at the various steps of the construction phase. 
   Extensive test-beam measurements have been performed with
   prototype and final detector
 modules, which have also allowed the validation
   of the simulation packages (e.g. GEANT4) used for instance
    to extrapolate the detector response from the test-beam  
    to the collider environment. Such detailed checks 
    and tests represent an unprecedented culture in our field.
        In addition, {\it in situ} commissioning and 
   calibrations after installation in the pits 
   will be needed to understand the experiments as a whole, 
   to account for the presence of e.g. upstream material and
   magnetic field, to cure long-range effects, etc. These calibrations
  will be based
  on cosmic muons, beam-halo muons and beam-gas events during the
  pre-collision phase (i.e. in the first half of 2007, 
  during the machine cool-down and single-beam 
  commissioning). Then, as
  soon as first collisions will be available, well-known physics samples 
  (e.g. $Z\to\ell\ell$ events, see tab.~\ref{tab_perf}) will be used.  
   
    As an example of the above procedure, the case of the
   ATLAS lead-liquid argon 
electromagnetic calorimeter\cite{largtdr}, for which
   the construction phase
   is completed, is discussed below.  
 
  One crucial performance issue for the LHC electromagnetic calorimeters
  is to provide a mass resolution of about 1\% in the hundred GeV
  range, needed to observe a possible $H\to\gamma\gamma$ signal as a 
  narrow peak on top of the huge $\gamma\gamma$ irreducible background.
   This requires a response uniformity, that is a total constant term
   of the energy resolution, of $\leq$0.7\% over the full calorimeter
   coverage ($|\eta|<2.5$). Achieving this goal is challenging, especially
   at the beginning, but is necessary for a fast discovery, and can 
   hopefully be accomplished in four steps:

\begin{figure}[t]
\begin{center}
\epsfig{file=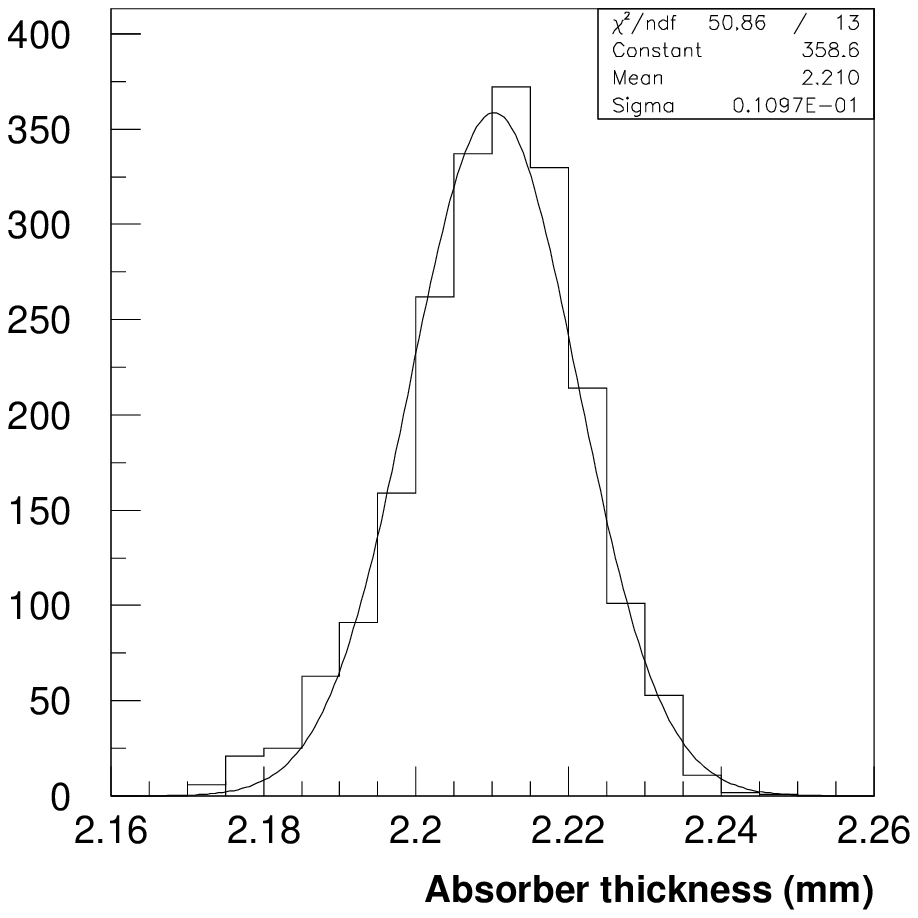,width=0.90\textwidth}\hspace{-6cm}
\epsfig{file=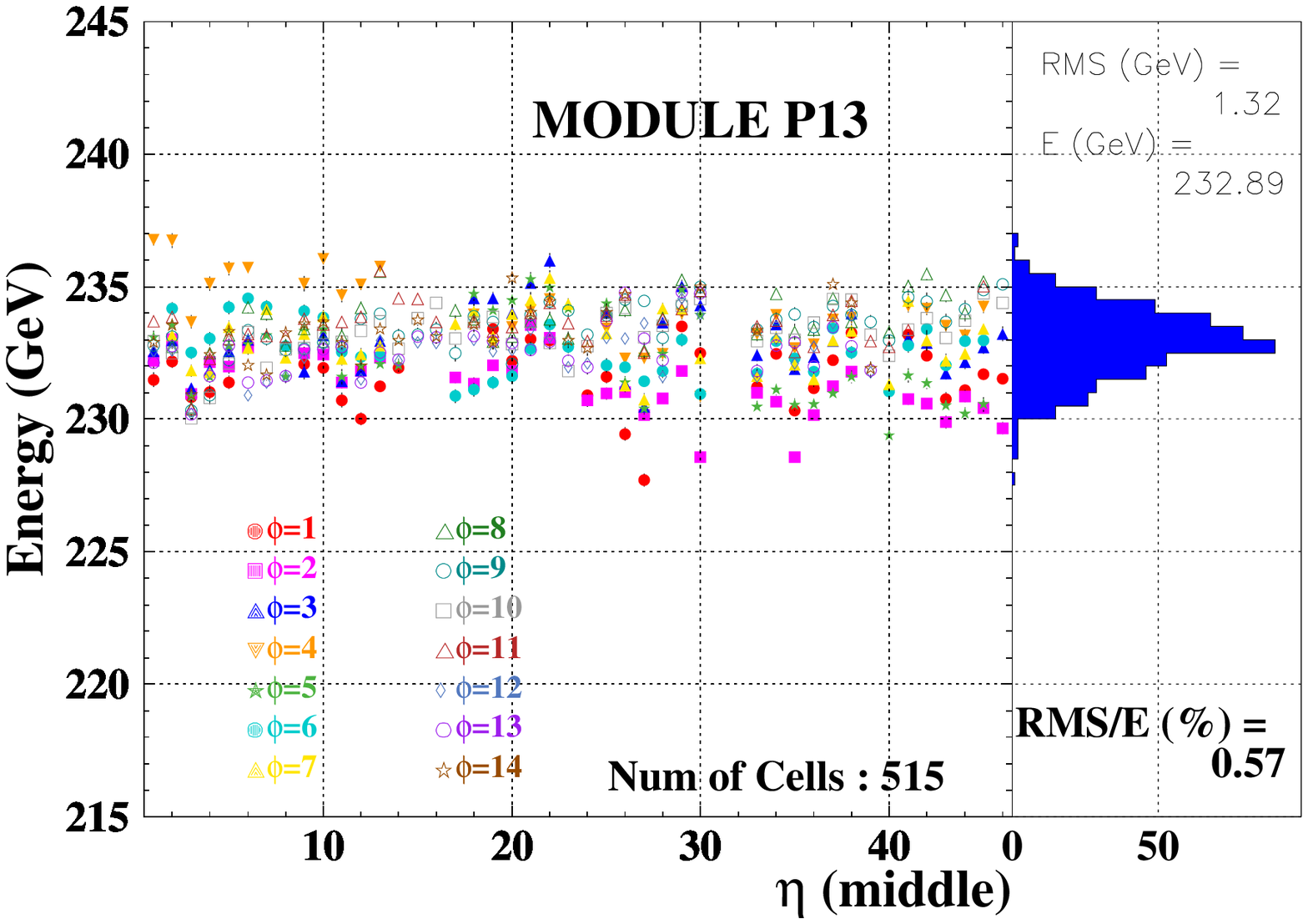,width=0.48\textwidth}
\caption{\it Left: Distribution of the thickness of the 2048 
   absorber plates (3~m long and 0.5~m wide) of the
  ATLAS barrel ECAL, as obtained from ultrasound measurements.
  The mean value of the distribution is 2.2~mm and the r.m.s.
  is 11~$\mu$m. Right: 
  Energy response of one module (of size
   $\Delta\eta\times\Delta\phi=1.4\times0.4$) of the ATLAS barrel ECAL,
  as a function of rapidity, as measured from a scan 
  with test-beam electrons. The various symbols 
  indicate different $\phi$ rows.\label{lead}}
\end{center}
\end{figure}

\begin{itemize}    
      
\item  {\underline {Construction quality}}. 
    Test-beam measurements performed with prototypes of the ATLAS ECAL
    in the early '90s showed 
   that a 1\% excess in the thickness of the lead plates produces
   a drop of the calorimeter response by 0.7\%.  Therefore, in order
    to keep the maximum response non-uniformity coming from the detector
   mechanics alone below 0.3\%, the thickness of the lead 
   plates must be uniform to about 0.5\%, i.e. $\sim$10~$\mu$m. This goal has
   been achieved, as shown in the left panel of fig.~\ref{lead}.  
   
\item {\underline {Test-beam measurements}}. 
    About 15\% of the final calorimeter modules
   have been exposed to electron beams, in order to verify
   the construction uniformity and to prepare  
   correction factors to the detector response. 
    The right panel in fig.~\ref{lead} shows the results
   of a position scan of one module performed
   with high-energy test-beam electrons. For all tested modules, 
   the response non-uniformity was found to be 
   about 1.5\% before correction, i.e. at the exit of the construction
   chain, and better than 0.7\% after calibration with test-beam data. 
   
\item  {\underline {Pre-collision phase}}. Before data taking starts, 
    the calorimeter calibration can be checked {\it in situ}
   with physics-like signals  by using  cosmic muons. 
     Table~\ref{tab_cosmics} shows the expected rates of
    cosmics in ATLAS\cite{boonekamp} as obtained from a full simulation
    of the detector inside the  underground cavern (including the
    overburden, the access shafts and the surface buildings).
     These results have also been validated by direct measurements of
     the cosmics flux in the pit made with a scintillator telescope. 
   It can be seen that rates between 0.5~Hz and 30~Hz are expected,
    depending on the requirements on the muon trajectory. 
     Therefore, in about three months
   of cosmics runs in 2007 during the machine cool-down and commissioning, 
   a few million events should be collected, a data sample
   large enough to catalog and fix several problems, gain
   operational experience, check the relative timing and position 
   of the various sub-detectors, etc., hopefully
   in a more relaxed environment than during the collision phase. 
   
    In particular, for what concerns the electromagnetic
    calorimeter, the signal-to-noise ratio 
    for muons is large enough ($S/N\sim$~7 from test-beam measurements) 
      that cosmic muons can be used 
   to check the calibration uniformity of the barrel calorimeter
   as a function of rapidity.
   The calorimeter is equipped with an electronics calibration
   system delivering pulses uniform to 0.25\%. However, the calibration
   signals and the physics signals do not have exactly the same shape,
    and the difference depends on the rapidity of a given
    calorimeter cell. This induces a
    non-uniformity of the ECAL response to incident particles
    as a function of $\eta$. Test-beam studies show that the expected 
    sample of cosmic muons is large enough to allow measurements of
    these effects down to the 0.5\% level.

\item {\underline {First collisions}}. 
  As soon as first collider data will be available,
  $Z\to~ee$ events, which are produced at the rate of $\sim$1~Hz
  at a luminosity of $10^{33}$~cm$^{-2}$~s$^{-1}$, will be used to
  correct long-range response non-uniformities from module to
  module, possible temperature effects, the impact of the upstream material, 
  etc. 
   Full simulation studies indicate that, since the
  calorimeter is already quite uniform on ``day 1" by construction and thanks
  to the previous steps, about $10^5$ $Z\to ee$ events should 
  be sufficient to achieve the goal overall constant term of 0.7\%.  
   In addition, this $Z\to ee$ sample should fix the absolute energy
   scale to about 0.5\%. 
    Therefore, after a few  weeks of data taking the
   ATLAS ECAL should in principle be fairly well calibrated. 
   
\end{itemize}

\begin{table}[t]
\centering
\caption{\it Expected rates of cosmic muons in ATLAS for various
 requirements on the muon trajectory, as obtained from a 
 full simulation of the detector inside the pit.} 
\vskip 0.1 in
\begin{tabular}{|l|c|c|}
\hline 
 topology & rate (Hz) &
comments \\
\hline\hline
through-going muons & $\sim$~25 & muons giving hits on top and bottom RPC's \\
                    &          & and in inner detector \\
close to interaction vertex & $\sim$~0.5 & muons passing within $|z|<60$~cm and $R<20$~cm\\
               &           &   from the interaction centre \\
useful for ECAL calibration & $\sim$~0.5 & muons  with $|z|<20$~cm, $E_{{\rm cell}}>100$~MeV \\
\hline
\end{tabular}
\label{tab_cosmics}
\end{table}

    As an academic exercise, one could consider 
    a very pessimistic (actually unrealistic...) scenario.  
   That is, ignoring the results and expectations discussed above, 
   one could assume that no corrections (neither based on test-beam data,
    nor using $Z\to ee$ events) will be applied. In this case,
    the intrinsic calorimeter constant term  would be given by the   
    uncorrected non-uniformity from detector construction
    (measured to be $\sim$1.5\%, as mentioned above), 
     to which another $\sim$1.5\% 
    from uncorrected material effects has to be added. This would give a  
    total constant term of the energy resolution of about 
    2\% instead of 0.7\%. As a consequence, 
     the significance of a $H\to\gamma\gamma$ signal
    would be reduced by about 30\%, and a factor
    1.7 more integrated luminosity would be needed to achieve
    the same sensitivity.

\section{How well will LHC physics and Monte Carlos 
 be known before data taking starts?}
While we cannot anticipate which new physics is waiting for us at the
LHC, we do know that there is plenty of SM processes to be
observed. In many cases, these processes offer themselves the potential
for important measurements (e.g. improved determinations of the $W$ and
top-quark masses, parton densities). More in general, they
will provide dangerous backgrounds to most signals of new physics. A
solid physics programme at the LHC will therefore require a robust
understanding of SM processes, and of QCD in particular. Significant
improvements have taken place in the past few years, as reviewed
in\cite{Mangano:2003ps} and shortly summarized here.

By far the cleanest process in $pp$ collisions, theoretically as well
as experimentally, is the production of
$W$ and $Z$ bosons. In addition to the full NNLO predictions for the
total cross sections, achieved long ago\cite{Hamberg:1990np}, NNLO
calculations for the experimentally more interesting rapidity
distributions have recently been
obtained\cite{Anastasiou:2003ds}, reducing the intrinsic
theoretical uncertainty for Drell-Yan cross sections to the level of
1-2\%. At this level of accuracy, EW effects start playing a role, as
recently evaluated in\cite{Baur:1998kt}, and a precise knowledge of the
parton densities (PDF) becomes essential. Progress in this field, in
addition to the availability of much more accurate data from
HERA\cite{hera}, has been driven by the development of formalisms
which allow a proper account of systematic uncertainties\cite{Martin:2002aw}.

The production of $t\bar{t}$ pairs, which at the Tevatron represents a
rather exotic signature, will become at the LHC a dangerous
background, with an inclusive rate of the order of 1~Hz. The cross
section is known from theory
with an accuracy of about 5\%\cite{Bonciani:1998vc},
enough to allow an indirect estimate of the top mass with an accuracy
of $\pm 2$ GeV (excluding experimental uncertainties). The ability to
precisely model the structure of the final states has
improved recently with the development of the MC@NLO code, where the
complete NLO parton-level matrix elements are consistently incorporated in
a full shower Monte Carlo (MC)\cite{Frixione:2002ik}. 
Also the description of bottom
quark production appears now to be under better theoretical control,
after improvements in the inputs of the calculations (fragmentation
functions and resummation of large logarithms) have led to excellent
agreement\cite{Cacciari:2003uh} with the most recent  results
from CDF\cite{Acosta:2004yw}. 

Complex multijet topologies can be described today more reliably,
thanks to recent advances in the calculation of multiparton final
states\cite{Caravaglios:1995cd}, their inclusion in parton level
codes\cite{Stelzer:1994ta,Mangano:2002ea}, and the development of
techniques to deal with the problem of properly merging with shower
MCs\cite{Catani:2001cc}. 
In addition, the well known and tested shower
MC codes which dominated the LEP and Tevatron era are being updated,
with inclusion of better algorithms for the development of the shower
or for the description of the underlying event\cite{Lonnblad:1992tz}.

 Validation of these new tools using Tevatron data
will be possible before the  LHC starts, but only the very large
statistics and the huge dynamic range of the LHC will allow complete
studies and proper tunings.

\section{Early physics goals and measurements\label{phys1}}

  Table~\ref{tab_stat} shows the data samples expected to be
 recorded by ATLAS and CMS for some example
 physics processes and for an integrated
 luminosity of 10~fb$^{-1}$. The trigger
 selection efficiency has been included. 
   Already over the first year (even days
 in some cases) of operation, huge event samples should
 be available from known SM processes, which will allow
 ATLAS and CMS to commission the detectors, the software and 
 the physics itself, and
 also from several new physics scenarios. We stress that
  this will be the case 
 even if the integrated luminosity collected during the
 first year were to be a factor of hundred smaller,
  i.e. $\sim$100~pb$^{-1}$.  
     
   In more detail, the following goals can be addressed
   with such data samples\footnote{It should be noted that the total 
   amount of data recorded by each experiment in one year of operation 
   corresponds to about 1 Petabyte, which represents
   an unprecedented challenge also for the LHC  
   computing and offline software.}:

\begin{itemize}

\item Commission and calibrate the detectors
 {\it in situ}, as already mentioned. 
   Understanding the trigger performance in as an unbiased way as possible, 
   with a combination of minimum-bias events, QCD jets
   collected with various thresholds, single and dilepton
   samples, is going to be one of the most challenging and 
   crucial steps at the beginning.  
  $Z\to\ell\ell$ is a gold-plated process for
 a large number of studies,  e.g. to set the absolute
 electron and muon scales in the ECAL and tracking detectors
 respectively,  whereas $t\overline{t}$ events can be used for
 instance to establish the absolute jet scale and to understand
 the $b$-tagging performance.  

\item Perform extensive measurements of the main SM physics
 processes, e.g. cross sections and event features for
 minimum-bias, QCD dijet, $W, Z, t\overline{t}$ production, etc. 
  These measurements will be compared to the predictions
  of the MC simulations, which  will already be quite
  constrained from theory and from studies at the Tevatron and HERA energies. 
  Typical initial precisions may be  10-20\% for 
  cross section measurements, and 5-7~GeV on the top-quark mass, and
  will likely be limited by systematic uncertainties after just a few weeks
  of data taking. 

\item Prepare the road to discoveries by measuring the backgrounds
 to possible new physics channels. Processes like $W/Z$+jets,
 QCD multijet production and $t\overline{t}$ are omnipresent 
 backgrounds for a large number of searches and need to
 be understood in all details. In addition, dedicated control
 samples can be used to measure specific backgrounds. For instance,  
 $t\overline{t}jj$ production, where
 the jets $j$ are tagged as light-quark jets, can be 
 used to gauge the irreducible $t\overline{t}b\overline{b}$ background
 to  the $t\overline{t}H\to t\overline{t}b\overline{b}$ channel. 

\end{itemize}

\begin{table}
\centering
\caption{\it For some  physics processes, the numbers of events expected
to be recorded by ATLAS and CMS for an integrated luminosity
of 10~fb$^{-1}$ per experiment.} 
\vskip 0.1 in
\begin{tabular}{|l|c|}
\hline 
 channel & recorded events per experiment for 10~fb$^{-1}$ \\
\hline\hline
$W\to\mu\nu$ & $7\times10^7$ \\

$Z\to\mu\mu$ & $1.1\times10^7$ \\

$t\overline{t}\to\mu+X$ & $0.08\times10^7$ \\

QCD jets $p_T>$150~GeV & $\sim10^7$ (assuming 10\% of trigger bandwidth) \\

minimum bias & $\sim10^7$  (assuming 10\% of trigger bandwidth) \\

$\tilde{g}\tilde{g}$,  m ($\tilde{g}$)$\sim$1~TeV & $10^3-10^4$ \\		
\hline
\end{tabular}
\label{tab_stat}
\end{table}

\begin{figure}[t]
\begin{center}
\epsfig{file=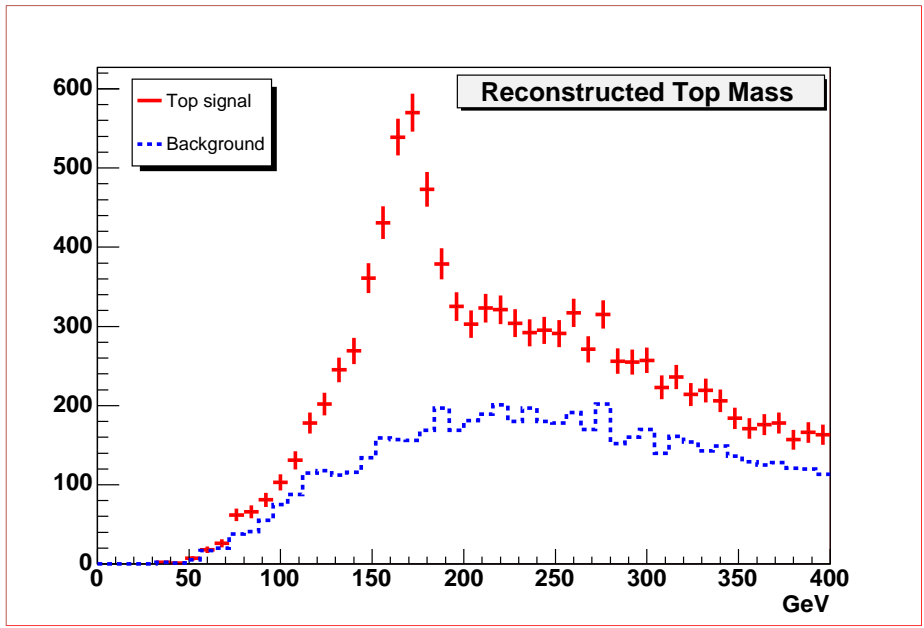,,width=0.8\textwidth}
%  hoffset=90 hscale=29 vscale=29 angle=0}
\caption   
{\it Three-jet invariant mass distribution for events selected as
described in the text, as obtained from a simulation of the
ATLAS detector. The dots with error bars 
show the expected signal from $t\overline{t}$ events plus the
background, the dashed line shows the $W+$4-jet background alone
 (ALPGEN Monte Carlo\cite{Mangano:2002ea}). 
  The number of events corresponds
 to an integrated luminosity of 150~pb$^{-1}$.\label{top_initial}}
\end{center}
\end{figure}

  As an example of initial measurement with limited detector
 performance, fig.~\ref{top_initial} shows the reconstructed
 top-quark signal in the gold-plated $t\overline{t}\to bjj\ b\ell\nu$ 
 semileptonic channel, 
  as obtained from a simulation of the ATLAS detector.  
  The event sample corresponds to an integrated luminosity of 150~pb$^{-1}$,
   which can be collected in less than one week of data taking at
   $L=10^{33}$~cm$^{-2}$~s$^{-1}$.  
   A very simple analysis was used to select these events, requiring an
  isolated electron or muon with $p_T>20$~GeV  and four
  and only four jets with $p_T>40$~GeV. The invariant mass
  of the three jets with the highest $p_T$ was then plotted. No kinematic
  fit was made, and no $b$-tagging of some of the jets was required, 
   assuming conservatively that the $b$-tagging performance would not
   have been well understood yet. Figure~\ref{top_initial}
   shows that, even under these over-pessimistic conditions, a clear top
   signal should be observed above the background after a few weeks
   of data taking (30~pb$^{-1}$ would be sufficient).
   In turn, this signal can be used for an early validation
   of the detector performance. For instance, if the top mass is wrong
   by several GeV, this would indicate a problem with the jet energy scale. 
    Furthermore, top events are an excellent sample to understand
    the $b$-tagging performance of ATLAS and CMS.   
    It should be noted that, unlike at the LHC, 
      at the Tevatron today the statistics of
    $t\overline{t}$ events is not sufficient to use these samples
    for detector calibration purposes.

\section{Early discoveries\label{phys2}} 

   Only after the three steps outlined in section~\ref{phys1} 
  will have been fully addressed can the LHC experiments
  hope to extract a convincing discovery signal from their data.          
    Three  examples of new physics 
   are discussed briefly below, ranked by increasing
   difficulty for discovery in the first year(s) of operation: 
   an easy case, namely a possible $Z'\to e^+ e^-$ signal,
   an intermediate case, SUSY, and a difficult
   case, a light Standard Model Higgs boson.

\subsection{$Z'\to e^+e^-$}   

  A particle of mass 1-2~TeV decaying into $e^+e^-$ pairs,
   such as a possible new gauge boson $Z'$,
 is probably the easiest object to discover at the LHC, for
 three main reasons. First, if the branching ratio into leptons is
 at least at the percent level as for the $Z$ boson, the expected
 number of events after all experimental cuts is relatively large, 
  e.g. about ten for an integrated luminosity as low as
  300~pb$^{-1}$ and a particle mass of 1.5~TeV.  
   Second, the dominant background,
   dilepton Drell-Yan production, is small in the
  TeV region, and even if it were to be a factor of two-three 
  larger than expected today (which is unlikely for such a theoretically 
  well-known process), it would still be negligible compared to the
  signal. Finally, the signal will be indisputable,
  since it will appear as a resonant peak on top
  of a smooth background, and not just as an overall excess
  in the total number of events.
    These expectations are not based on ultimate
   detector performance, since they hold
   also if the calorimeter response
   is understood to a conservative level of a few percent.

\subsection{Supersymmetry}    
    
\begin{figure}[t]
\begin{center}
\epsfig{file=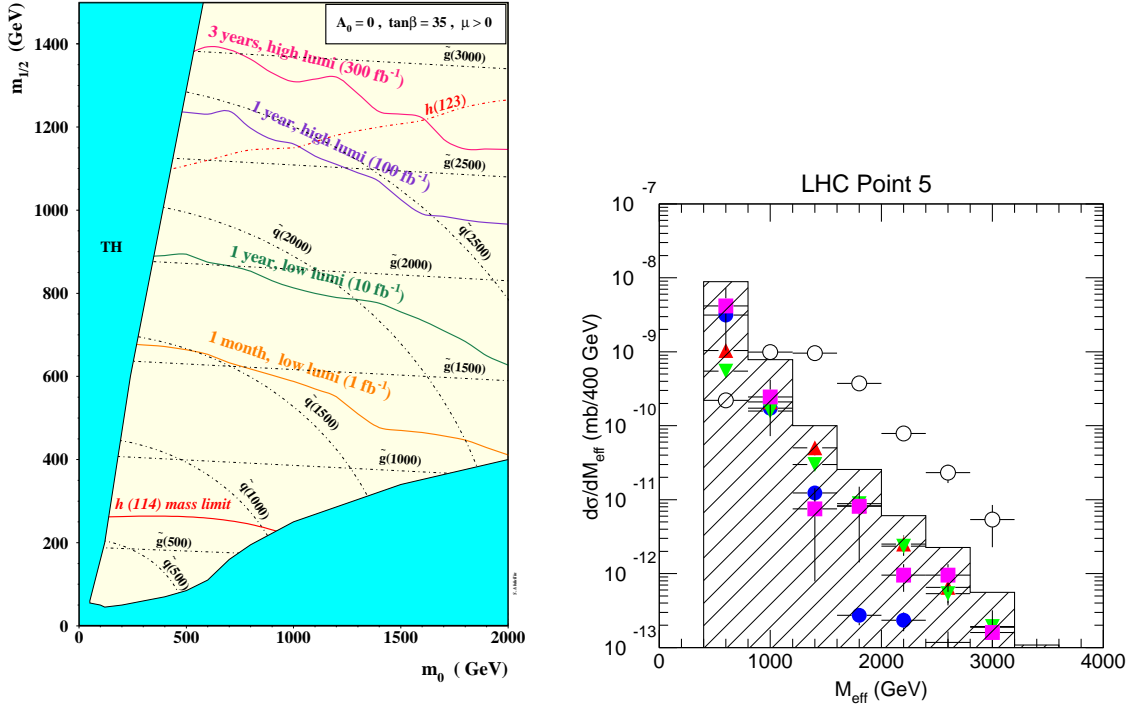,width=0.48\textwidth}\hfill
\epsfig{file=susy_atlas.epsi,width=0.48\textwidth}
\caption{\it Left: The CMS discovery potential\cite{cmstp} 
for squarks and gluinos
in mSUGRA models, parametrized in terms of the 
universal scalar mass $m_0$ and universal 
gaugino mass $m_{1/2}$,  as a function
of integrated luminosity. Squark and gluino mass isolines
are shown as dot-dashed lines (masses are given in GeV).
 Right: The expected distribution of the effective mass (see text) for
 the SUSY signal at ``Point~5"\cite{phystdr} of the mSUGRA parameter 
 space (open circles), as obtained from a simulation of the ATLAS
 detector. The histogram shows the total SM background, which
 includes $t\overline{t}$  (solid circles), 
 $W$+jets (triangles), $Z$+jets (downward triangles), and QCD jets
  (squares). 
 \label{fig_susy}} 
\end{center}
\end{figure}

   Extracting a convincing signal of SUSY in the early phases 
   of the LHC operation is not as straightforward as for the
   previous case, since good calibration of the detectors and
  detailed understanding of the numerous backgrounds are
  required. As soon as these
  two pre-requisites are satisfied, observation of a
  SUSY signal should be relatively easy and fast.   
   This is because of the huge
   production cross sections, and hence event rates,
  even for squark and gluino masses as large as  $\sim$1~TeV (see 
  tab.~\ref{tab_stat}), and the clear signature of such events
  in most scenarios. Therefore,
   by looking for final states containing several high-$p_T$ jets
   and large \met, which is the most
   powerful and model-independent signature if R-parity is
   conserved, the LHC experiments should be able to discover
  squarks and gluinos up to masses of $\sim$1.5~TeV in only
  one month of data taking at $L=10^{33}$~cm$^{-2}$~s$^{-1}$,
  as shown in the left panel of fig.~\ref{fig_susy}.

   Although detailed   
  measurements of the SUSY particle masses will likely 
  take several years, it should nevertheless be
  possible to obtain a first determination of the SUSY 
  mass scale quickly after discovery. This is illustrated
  in the right panel of fig.~\ref{fig_susy}, which shows the 
  striking SUSY signal on top of the SM background,
   expected at a point in the minimal SUGRA parameter space
   where squark and gluino masses are about 700~GeV. 
   The plotted variable, called ``effective mass" ($M_{{\rm eff}}$), is 
   defined as the scalar sum
  of the event \met\ and of the transverse
  energies of the four highest $p_T$ jets, and thus
  reflects the ``heaviness" of the particles produced in the final state.
    More
   precisely, the position of the peak of the $M_{{\rm eff}}$ signal 
    distribution 
   (see fig.~\ref{fig_susy})  moves to larger/smaller values with 
   increasing/decreasing squark and gluino masses. 
   Therefore a measurement of the signal peak position should provide
   a first  fast determination of the mass scale of SUSY.
  The expected precision is about 20\% for an integrated luminosity
  of 10~fb$^{-1}$, at least in minimal models like mSUGRA. 
     
   A crucial detector performance issue for an early SUSY
  discovery is a reliable reconstruction of the event
  \met, 
   which is {\it a priori} prone to contamination
  from several instrumental effects (calorimeter non-linearities, 
  cracks in the detector, etc.). Final states with non-genuine
  \met\ can be rejected by requiring the event primary
   vertex to be located close to the interaction centre (which also
   helps to suppress the background from cosmic and beam-halo muons), 
   no jets pointing to detector cracks, and that the missing $p_T$ vector 
   is not aligned with any jet.  
    The calorimeter response linearity can
   be understood to a large extent 
   by using ``calibration" samples like $Z (\to\ell\ell)$+jet events
    (with $\ell=e,\mu$), where the lepton pair and the jet
    are back-to-back in the transverse plane, 
   so that the well-measured $p_T$ of the lepton pair can be used 
   to calibrate the jet $p_T$ scale over a large dynamic range.  

   Concerning the physics backgrounds
    (e.g. $Z\to\nu\overline{\nu}$+jets, $t\overline{t}$ production,
    QCD multijet events), most of them can be measured by using
    control samples. For instance, $Z\to\ell\ell$+jet production
    provides a normalization of the $Z\to\nu\overline{\nu}$+jets
    background.  More difficult to handle is the residual background
    from QCD multijet events with fake \met\ produced by the
    above-mentioned instrumental effects.  The technique used at the
    Tevatron consists of normalizing the Monte Carlo simulation to the
    data in the (signal-free) region at low \met, and then use
    the Monte Carlo to predict the background in the (potentially
    signal-rich) region at large \met.

\begin{figure}[t]
\begin{center}
\epsfig{file=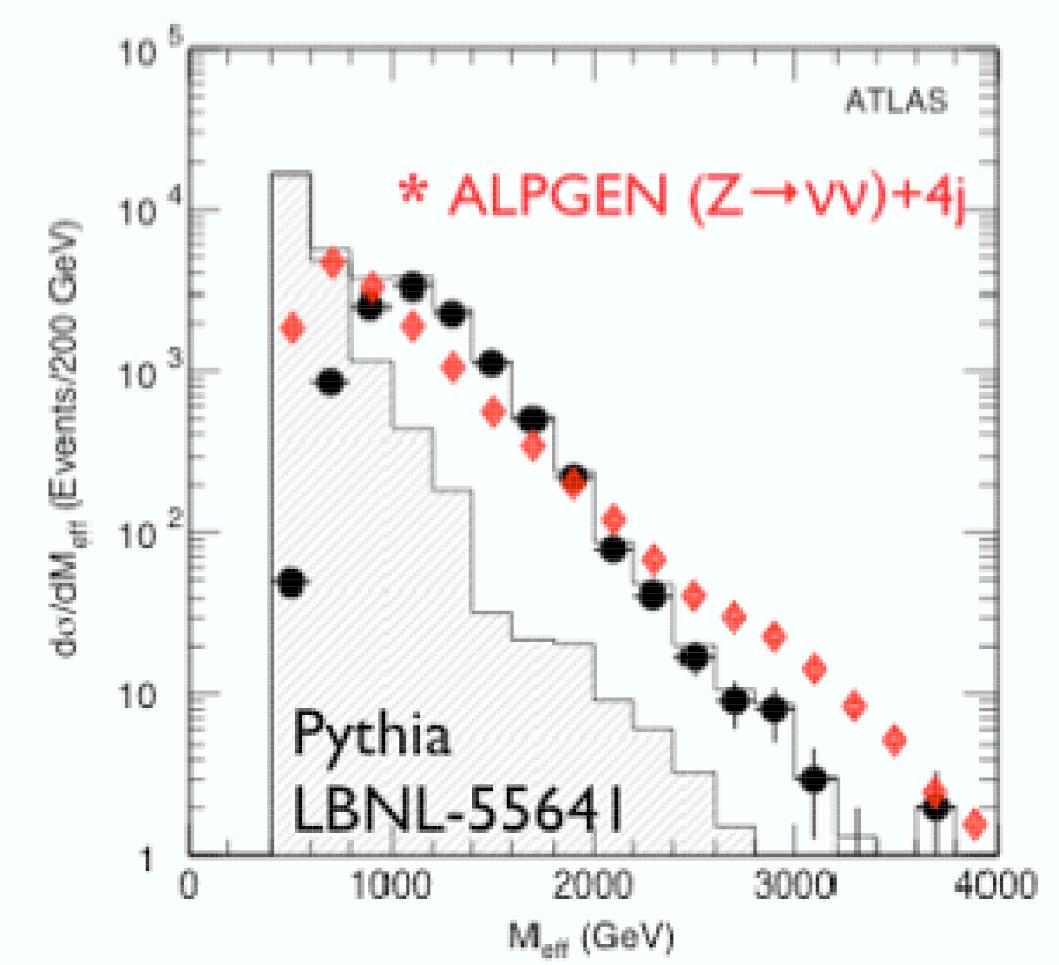,width=0.8\textwidth}
\caption{\it $M_{\mathrm{eff}}$ distributions for a potential SUSY
 signal (histogram), separated into signal (dark
 points) and the background prediction from shower MC (shaded
 histogram), compared to the $Z(\to \nu\bar{\nu})$+4jet background evaluated
 with exact matrix elements (grey diamonds). \label{fig:zjets}}	 
\end{center}
\end{figure}

A crucial element in the ability to calibrate these backgrounds using
the theoretical MC predictions to extrapolate from the signal-free
to the signal-rich regions is the reliability of the MC themselves. As
mentioned earlier, their level of accuracy and their capability to
describe complex final states, such as the multijet topology typical of new
phenomena like SUSY, have improved significantly over the past few
years\cite{Mangano:2003ps}. In some cases, the predictions obtained
with the new tools are very different from those derived in the
past. In particular, the description of multijet final states, which
until the recent past could only be achieved in a rather approximate
way with shower MCs, is now performed starting from exact
matrix-element calculations of the multiparton emission
amplitudes. This typically results in higher production rates,
increasing therefore the difficulty of extracting in a robust way the
signals of new physics from the QCD backgrounds. An example of this is
shown in fig.~\ref{fig:zjets}: the diamond plot represents the
matrix-element 
prediction\cite{Mangano:2002ea} of the
$Z(\to \nu\bar{\nu})$+4jet background to a possible multijet+\met\
SUSY signal, compared to estimates (the grey histogram, which also
includes the contribution of \met-mismeasurement in pure multijet events) 
obtained in
the past with standard shower MC simulations. Not only is the rate
larger than previously expected, but the shape of the distribution is
different, and much closer to that of the signal itself. A calibration
of the absolute rate using $(Z\to \ell^+\ell^-)$+4jet data is still
possible where the statistics allow (up to $M_{\mathrm{eff}}\sim 1-2$~TeV), but
a validation of the MCs is clearly required to ensure a robust
extrapolation to the highest values of   $M_{\mathrm{eff}}$. 

\subsection{Standard Model Higgs boson}

\begin{figure}[t]
\begin{center}
\epsfig{file=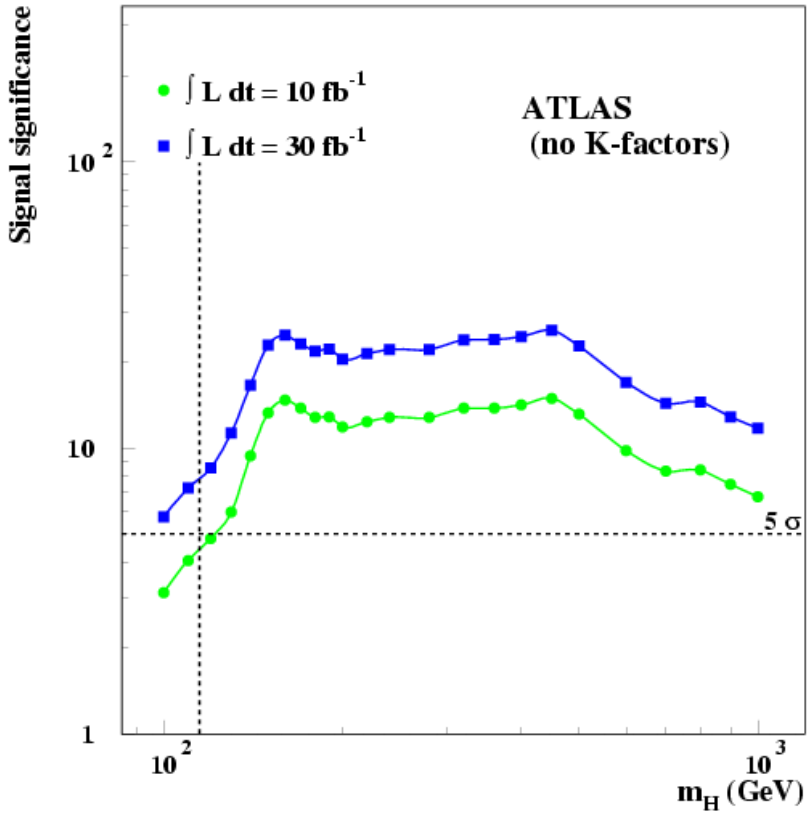,width=0.8\textwidth}
\caption{\it The expected signal significance 
 for a SM Higgs boson in ATLAS 
 as a function of mass,  for 
 integrated luminosities of 10~fb$^{-1}$ (dots) and 30~fb$^{-1}$ 
 (squares). The vertical line shows the  lower limit from searches at LEP.
  The horizontal line indicates the minimum significance ($5\sigma$)
  needed for discovery.\label{atlas_new}}
\end{center}
\end{figure}

   The possibility of discovering a SM Higgs boson at the LHC 
  during the first year(s) of operation depends very much on the Higgs mass, 
   as shown in  fig.~\ref{atlas_new}.  
    If the Higgs mass is larger than 180~GeV, discovery may be relatively
    easy thanks to the gold-plated $H\to 4\ell$ channel, which is essentially
    background-free. The main requirement in this case is an 
    integrated luminosity of at least 5-10~fb$^{-1}$, since the signal 
    has a cross section of only a few fb.
    
     The low-mass region close to the LEP limit is much more difficult.
    The expected sensitivity for a Higgs mass of 115~GeV and for
    the first good  (i.e. collected with well-calibrated
    detectors)  10~fb$^{-1}$ is summarized in tab.~\ref{tab_higgs}. 
   The total significance of about 4$\sigma$ per experiment
   ($4^{+2.2}_{-1.3}~\sigma$ including the expected systematic uncertainties)    
  is more or less equally 
   shared among three channels: $H\to\gamma\gamma$, 
   $t\overline{t}H$ production with
   $H\to b\overline{b}$, and Higgs production in vector-boson fusion
   followed by $H\to\tau\tau$. 
\begin{table}[t] 
\centering  
\caption{\it For a Higgs boson mass of 115~GeV and  an integrated luminosity
 of 10~fb$^{-1}$, the expected
 numbers of signal ($S$) and background  ($B$) events after all cuts
 and signal significances ($S/\sqrt{B}$) in ATLAS for the three
 dominant channels.}
\vskip 0.1 in  
\begin{tabular}{|l|c|c|c|}
\hline
  & $H\to\gamma\gamma$ & $t\overline{t}H\to t\overline{t} b\overline{b}$ 
  & $qqH\to qq\tau\tau\to\ell+X$ \\ 
\hline\hline
$S$ & 130 & 15 & $\sim10$ \\
$B$ & 4300 & 45 & $\sim10$ \\
$S/\sqrt{B}$ & 2.0 & 2.2 & $\sim 2.7$ \\
\hline
\end{tabular}
\label{tab_higgs}
\end{table}     
     A conservative approach has been adopted in deriving these
     results. For instance, very simple cut-based analyses have been used, 
    and higher-order corrections to 
   the Higgs production cross sections (the so-called K-factors),
    which are expected to increase for example the $gg\to H\to\gamma\gamma$
  rate by a factor of about two compared to leading order, have not been included.          
    Nevertheless, it will not be easy to extract
   a convincing signal with only 10~fb$^{-1}$, 
   because the significances of the individual channels are small, and
    because an excellent knowledge of the backgrounds and close-to-optimal
    detector performances are required, as discussed below.    
    Therefore, the contribution of both experiments,  and the 
    observation of possibly all three channels, will be crucial 
  for an early discovery. 
  
   The channels listed in
  tab.~\ref{tab_higgs} are complementary. They are characterized
  by different Higgs production mechanisms and decay modes, and
  therefore by different backgrounds and 
  different detector requirements. 
  Good uniformity of the electromagnetic calorimeters
 is crucial for the $H\to\gamma\gamma$ channel, as already mentioned.  
   Powerful $b$-tagging is the key performance 
 issue for the $t\overline tH$ channel, since there are four $b$-jets 
 in the final state which all need to be tagged in order 
 to reduce the background. 
  Efficient and precise jet reconstruction over ten rapidity units ($|\eta|<5$)
is needed for the $H\to\tau\tau$ channel, since tagging 
the two forward jets accompanying the Higgs boson and vetoing additional
jet activity in the central region of the detector 
are necessary tools to defeat the background. 
  Finally, all three channels demand
   relatively low trigger thresholds (at the
 level of 20-30~GeV on the lepton or photon $p_T$), and a control of
   the backgrounds to a few percent. These requirements are 
  especially challenging during the first year(s) of operation.

\section{Conclusions}

  The LHC offers the potential for very interesting
 physics and major discoveries right from the beginning. 
 We note that for some standard physics processes, a single day 
 of data taking at $L=10^{33}$~cm$^{-2}$~s$^{-1}$ corresponds, in terms
 of event statistics, to ten years of operation at previous machines. 
  SUSY may be discovered
 quickly, a light Higgs boson will be much more difficult to observe,
 unexpected scenarios and surprises may also be round the corner
 at an unprecedented collider exploring a completely new territory. 
 
 The machine luminosity performance will be the crucial
issue at the beginning. Hopefully, an instantaneous
luminosity of  up to $L\sim10^{33}$~cm$^{-2}$~s$^{-1}$, and 
an integrated luminosity of a few fb$^{-1}$ per experiment, 
 can be achieved by the end of 2008, as estimated by the 
 accelerator team. 

 Concerning the experiments, a lot of emphasis has been
given to quality checks in the various phases of the construction and
to tests with beams. The results indicate that the detectors ``as built"
should give a good starting-point performance already on ``day 1". 
   However, a lot of data and time will be needed to commission the
 detectors, the triggers and the software {\it in situ}, 
  to reach the performance required to address serious physics studies,
  to understand standard physics and the Monte Carlo tools 
   at $\sqrt{s}$=14~TeV, and to measure 
  the backgrounds to possible new physics processes. 
  
    The next challenge is therefore an efficient and timely 
   detector commissioning, from cosmics runs to first collisions, 
  where the experiments try to learn and fix as much as 
  possible as early as possible. 
   In parallel, efforts to improve and tune the Monte Carlo
  generators, based on theoretical developments as well
  as on comparisons with data from past and present experiments, 
   should be pursued with vigor. 
  Indeed, both activities will be crucial to reach quickly the discovery
  phase, and to extract  convincing signals of new physics in the
  first year(s) of operation.

\end{document}